\documentclass{PoS}

\title{Dark matter line searches towards dwarf galaxies with H.E.S.S.}

\ShortTitle{DM Line Search with H.E.S.S.}

\author{\speaker{Louise Oakes}\\
        Humboldt University Berlin\\
        E-mail: \email{loakes@physik.hu-berlin.de}}
\author{Aion  Viana\\
        MPI Heidelberg\\
        E-mail: \email{aion.viana@mpi-hd.mpg.de}}
\author{Emmanuel Moulin\\
        CEA\\
        E-mail: \email{emmanuel.moulin@cea.fr}}
\author{Lucia Rinchiuso\\
        CEA\\
        E-mail: \email{lucia.rinchiuso@cea.fr}}
\author{{Ullrich Schwanke}\\
        Humboldt University Berlin\\
        E-mail: \email{schwanke@physik.hu-berlin.de}}
\author{{for the HESS Collaboration, and }}
\author{Marco Cirelli\\
        CNRS\\
        E-mail: \email{marco.cirelli@gmail.com}}
\author{Paolo Panci\\
        CERN\\
        E-mail: \email{paolo.panci@cern.ch}}
\author{Filippo Sala\\
        CNRS\\
        E-mail: \email{filo.sala@gmail.com}}
\author{Joseph Silk\\
        IAP, JHU, Oxford, CEA\\
        E-mail: \email{j.silk1@physics.ox.ac.uk}}
\author{Marco Taoso\\
        CEA\\
        E-mail: \email{m.taoso@csic.es}}

\abstract{High energy $\gamma$-rays are powerful probes in the search for annihilations of dark matter (DM) particles in dense environments. In several DM particle models their annihilation produces characteristic features such as lines, bumps or cut-offs in their energy spectrum. The High Energy Stereoscopic System (H.E.S.S.) of imaging atmospheric Cherenkov telescopes is perfectly suited to search for such features from multi-TeV mass DM particles. The Dwarf Spheroidal Galaxies (dSphs) of the Local Group are the most common satellites of the Milky Way and assumed to be gravitationally bound dominantly by DM, with up to O(10$^3$) times more mass in DM than in visible matter. Over the past decade, several observational campaigns on dwarf satellite galaxies were launched by H.E.S.S. amounting to more than 140 hours of exposure in total. The observations are reviewed here. In the absence of clear signals, the expected spectral and spatial morphologies of signal and background are used to derive constraints on the DM particle annihilation cross-section for particle models producing line-like signals. The combination of the data of all the dwarf galaxies allows a significant improvement in the HESS sensitivity.
}

\FullConference{35th International Cosmic Ray Conference --- ICRC2017\\
		10--20 July, 2017\\
		Bexco, Busan, Korea}

\begin{document}

\section{Introduction}
Predicted to be some of the most Dark Matter (DM) dominated objects in the universe, Dwarf Spheroidal Galaxies (dSphs) are a promising target for DM searches due to their expected large signal to noise ratio. Indirect DM searches with Cherenkov telescopes such as those of the High Energy Stereoscopic System (H.E.S.S.) experiment~\cite{HESS1} can detect the presence of DM by observing $\gamma$-rays originating from the self-annihilation of DM particles such as Weakly Interacting Massive Particles (WIMPs).  A clear signal for WIMP annihilations would be the presence of a peak in the energy distribution of observed photons at the WIMP mass - known as a gamma-ray line, as there is no background astrophysical process that would give a similar effect. 
Previous H.E.S.S. publications have shown limits for a continuum DM signal from dwarf galaxies~\cite{HESS2} and line-like signals from the Galactic Centre (GC)~\cite{GC}. The search for gamma-ray lines from dwarf galaxies is subject to different astrophysical uncertainties than the GC, as recently emphasised in ~\cite{line1}. Here the combined constraints for a DM line signal from four dwarf galaxies observed by H.E.S.S. are presented. 

\section{The H.E.S.S. experiment}
The H.E.S.S. observatory is located at an elevation of 1800 m in the Khomas Highland of Namibia. The experiment is composed of four imaging atmospheric Cherenkov telescopes of 13 m diameter each with a mirror area of 108 m$^2$ and a larger central telescope with 28 m diameter and a surface area of 614 m$^2$. The data used in this analysis were taken during the H.E.S.S. I phase, with only the 4 smaller telescopes in operation.      
 
\section{Analysis}
\subsection{Observations and sources}
Data were collected during observations from 2006-2012 for four dwarf galaxies: Fornax dSph, Coma Berenices dSph, Sculptor dSph and Carina dSph. Observations were carried out with the source offset from the centre of the field of view, such that the gamma-ray-like cosmic-ray background flux can be simultaneously measured.  This mode of observation is called \textit{wobble mode}. The datasets used for the analysis include only the observations that meet standard quality control criteria~\cite{qc} based on the weather and data acquisition conditions.
\subsection{Data analysis\label{secdat}}
A \textit{model analysis}~\cite{model} chain was used for the selection of signal events and background suppression. This method is based on an accurate pixel per pixel comparison of the observed intensity with a pre-calculated semi-analytical model of showers. All analysis results were crosschecked using an independent calibration and analysis chain~\cite{ImPACT}, which gave compatible results.
Table~\ref{tab1} shows results of the analysis for each dwarf galaxy.
\begin{table}[!htb]
\centering
\begin{tabular}{l|ccccc}
\hline\hline
Source name & livetime (h) & nOn & NOff & $\alpha$ & significance($\sigma$) \\
\hline
Fornax &2.6 &192 &4943 &21.6 &-2.4 \\
Coma Berenices &10.0 &1513 &4191 &2.8 &0.4 \\
Sculptor &10.1  &1176 &11141 &8.91 &-2.0 \\
Carina &20.0 &1515 &11865 &8.02 &0.9 \\
\hline\hline
\end{tabular}
\caption{Summary of data analysis results for each dSph: acceptance corrected livetime,  number of events in signal region (ON-region), number of events in control region (OFF-region), exposure ratio and excess significance. \label{tab1} }
\end{table}

\subsection{Analysis technique}
The DM annihilation flux depends on two factors: the particle physics flux, $\Phi^{PP}$, which is defined by the underlying DM model, and the astrophysical factor, known as the $J$-factor which is dependent on the DM density distribution of the source. This can be written as:
\begin{equation}
\frac{d\Phi_{\gamma}}{dE_{\gamma}}(E_{\gamma},\Delta\Omega)=\Phi^{PP}(E_{\gamma})\times J(\Delta \Omega)
\end{equation}
where the particle physics factor, assuming self-conjugated DM particles, is:
\begin{equation}
\Phi^{PP} = \frac{1}{8\pi M^2_{\chi}} <\sigma v> \frac{dN}{dE_{\gamma}}
\end{equation}
and the astrophysical factor:
\begin{equation}
J(\Delta\Omega)=\int_{\Delta\Omega}\int_{l.o.s.}\textrm{ds} \textrm{d}\Omega \rho^2(r(s,\theta)).
\end{equation}
In the preceeding equations, $ <\sigma v>$ is the thermally-averaged velocity-weighted WIMP annihilation cross section,  $\frac{dN}{dE_{\gamma}}$ is the differential gamma-ray yield per annihilation and $J(\Delta\Omega)$ is the integral of $\rho^2$, the DM density squared, over $\Delta\Omega$, the solid angle, along the line of sight (l.o.s.).  The $J$-factors used in this analysis for each dSph are based on the results reported in~\cite{Jfac}.

The analysis assumes a monochromatic line-like WIMP annihilation signal with no expected gamma-ray background for the dSphs. The gamma-ray line has to be convolved by the energy resolution of H.E.S.S., which results in an expected signal with a Gaussian shape with width $\sigma_{\rm E} \sim$10\%-15\% for energies above 200 GeV. A 2D likelihood function binned in energy and spatial coordinates is used to calculate limits on the DM flux and the WIMP annihilation cross section, taking advantage of the spatial and spectral differences between the expected DM signal and measured residual cosmic-ray background. The spatial regions of interest (RoIs) are defined as circular concentric regions of 0.1$^{\circ}$ width each, and centered at the dSph galaxy. Due to the small angular extension of dwarf galaxies, for Sculptor, Carina and Fornax, only two spatial bins (from 0$^{\circ}$ to 0.2$^{\circ}$ radius), and for Coma Berenices three spatial bins (from 0$^{\circ}$ to 0.3$^{\circ}$ radius) are used. 

The likelihood is calculated for each considered DM particle mass, $M_{\chi}$, as a product over the spatial RoIs (bins i) and the energy bins (bins j).  The 2D binned spatial and spectral likelihood formula is composed of a Poisson ``on'' term (first term in Eqn.~\ref{eqn1}) and a Poisson ``off'' term (second term in Eqn.~\ref{eqn1}):
\begin{equation}
\mathcal{L}_{ij}(N_{sig},N_{bkg}| N_{ON},N_{OFF}) = {\frac{(N_{sig,i,j}+N_{bkg,i,j})^{N_{ON,i,j}}}{N_{ON,i,j}!}}e^{-(N_{sig,i,j}+N_{bkg,i,j})}\times \frac{(\alpha_{i,j}N_{bkg,i,j})^{N_{OFF,i,j}}}{N_{OFF,i,j}!}e^{-(\alpha_{i,j}N_{bkg,i,j})}
\label{eqn1}
\end{equation}
where the analysis variables are as defined in Section~\ref{secdat}, $N_{sig,i,j}$ is the number of predicted signal events in the bin $i,j$ and $N_{bkg,i,j}$ the number of expected background events, and $\alpha$ is the exposure ratio of the ON and OFF regions.   
\section{Results}
\subsection{Significance}
\begin{figure}[!h]
\centering
\includegraphics[width=5cm]{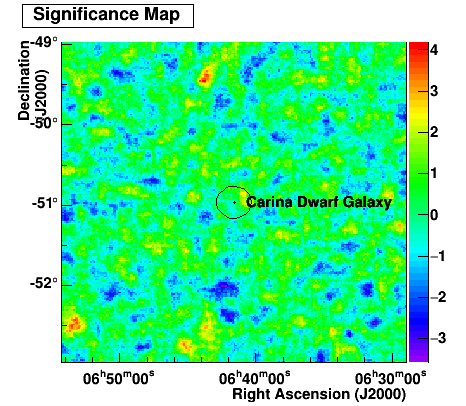}\includegraphics[width=5cm]{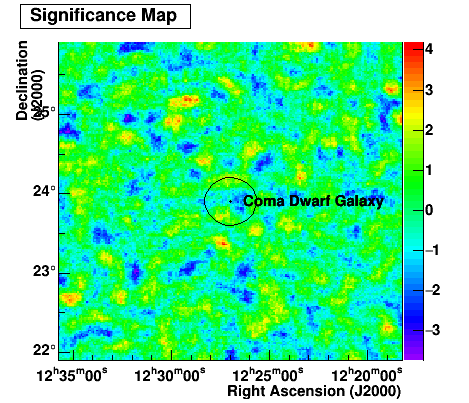}\\\includegraphics[width=5cm]{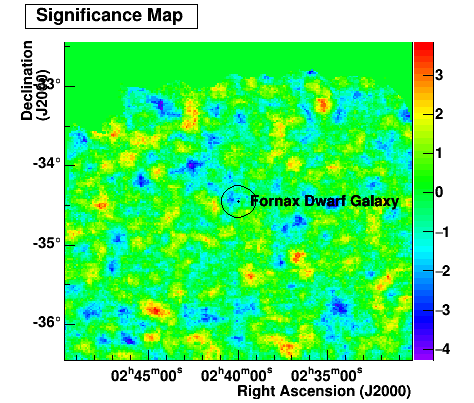}\includegraphics[width=5cm]{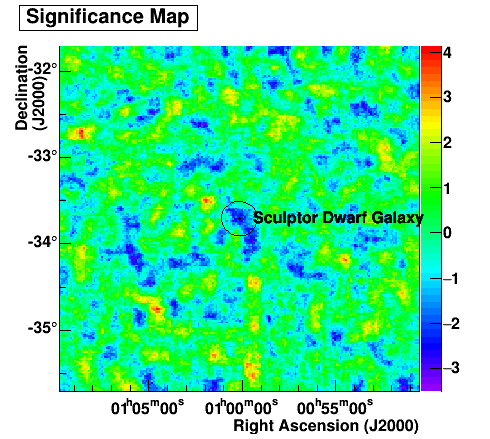}
\caption{Significance skymaps for the four studied dSphs: (top left) Carina, (top right) Coma Berenices, (bottom left) Fornax, (bottom right) Sculptor  \label{fig1}}
\end{figure}

\begin{figure}[!h]
\centering
\includegraphics[width=5cm]{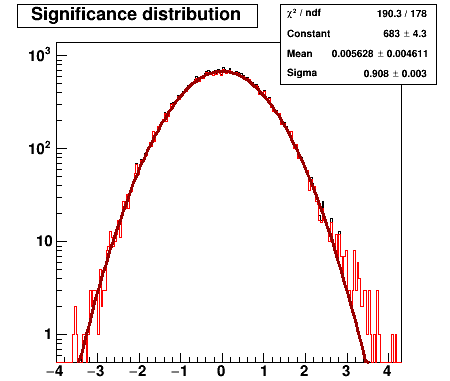}\includegraphics[width=5cm]{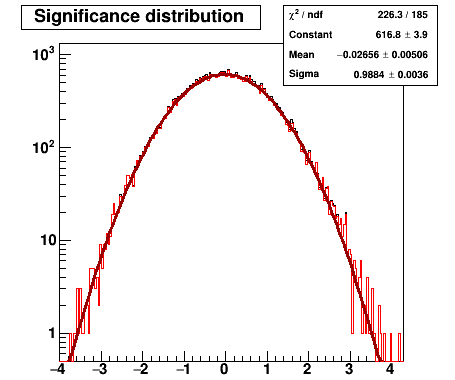}\\\includegraphics[width=5cm]{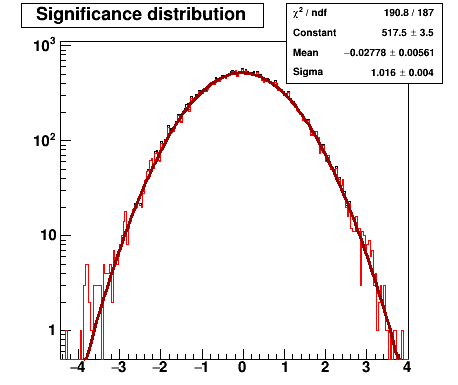}\includegraphics[width=5cm]{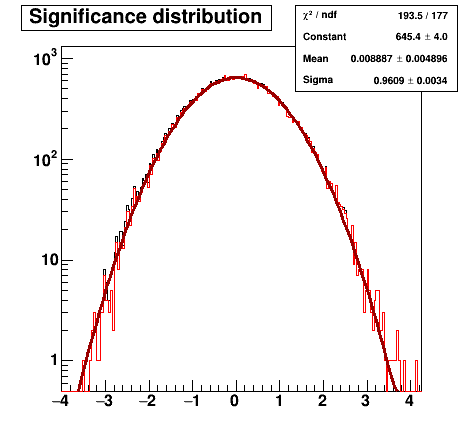}
\caption{Significance distributions over the field of view, fitted with a Gaussian function for the four studied dSphs: (top left) Carina, (top right) Coma Berenices, (bottom left) Fornax, (bottom right) Sculptor  \label{fig2}}
\end{figure}

There is no significant deviation from the expected background observed at the locations of the four dSphs. Figure~\ref{fig1} shows the significance maps for each dSph and in Figure~\ref{fig2} the significance distributions across the field of view are shown. The fitted significance distributions are compatible with a Gaussian with a centre of zero.

\subsection{Limits on DM flux}
\begin{figure}[!h] 
\centering
\includegraphics[width=8cm]{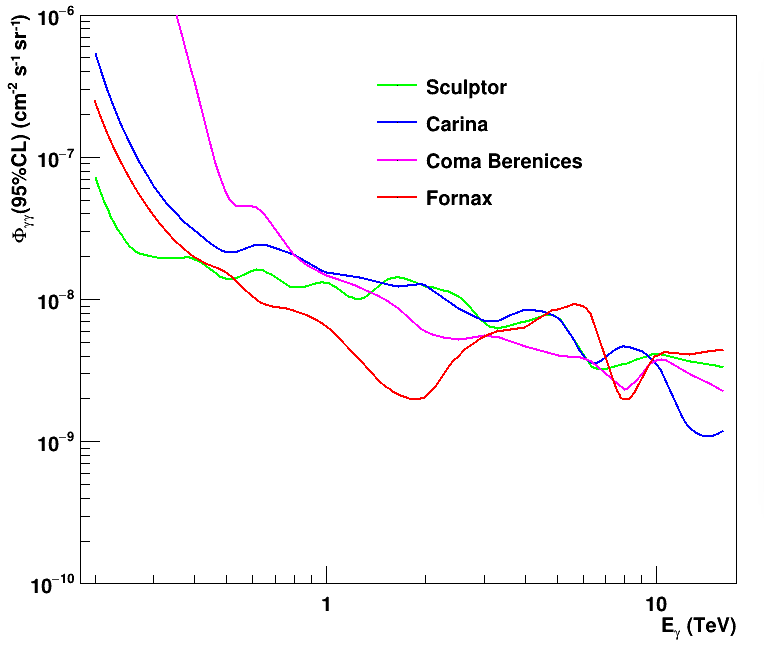}
\caption{ 95\% upper limits on the integrated flux as a function of photon energy $E_{\gamma}$ for Fornax, Carina, Coma Berenices and Sculptor dwarf galaxies. \label{figflux}}
\end{figure}

The calculated upper limits on the integrated flux in the range $\pm 5 \sigma_{E}$ of $M_{\chi}$ at the 95\% confidence level (CL) are shown in Figure~\ref{figflux} as a function of $M_{\chi}$. The differences in flux limits from the four dwarf galaxies arise from the various observation conditions and number of hours of livetime for each galaxy.
\subsection{Constraints on the WIMP cross-section}

\begin{figure}[!h] 
\centering
\includegraphics[width=8cm]{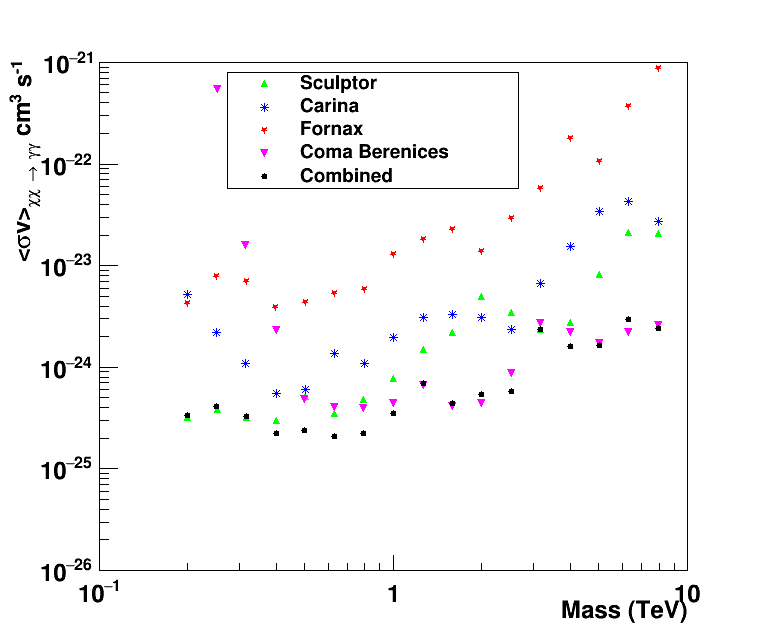}
\caption{95\% exclusion limits on the velocity weighted WIMP self-annihilation cross-section as a function of  $M_{\chi}$ for Fornax, Carina, Coma Berenices and Sculptor dwarf galaxies.  \label{figsigv}}
\end{figure}

Figure \ref{figsigv} shows the 95\% CL constraints on the velocity weighted WIMP self-annihilation cross-section ($<\sigma v>$) versus $M_{\chi}$ for the individual dSphs as well as a combined result for the four galaxies.  A limit of $~3\times 10^{-25} cm^3 s^{-1}$ is reached in the mass range of 400 GeV to 1 TeV. The combination of all the galaxies allows improvement in the constraints by up to a factor of 2 around 500 GeV with respect to individual galaxies.
\section{Conclusions and outlook} 
The initial results of the first DM line search towards Dwarf Galaxies with H.E.S.S. have been presented, showing upper limits on the DM flux and the velocity weighted WIMP self-annihilation cross-section for four dSphs and result for the combination of the constraints on $<\sigma v>$. This is the first time a combination of several dwarf galaxies is used to search for gamma-ray lines. Work is in progress to produce limits from H.E.S.S. data for more complex line-like DM signals~\cite{linemodels,linmod2}. Future plans include collecting further data on newly discovered dwarf galaxies~\cite{des} and adding these to the combined analysis. Beyond this work, the sensitivity of indirect DM searches~\cite{cta,cta2} including searches for DM line signatures will increase with the start of the new Cherenkov Telescope Array (CTA) which will have a factor of 2 better energy resolution and be able to detect gamma-rays down to significantly lower energies.

\end{document}